# Thermodynamic Model of Liquid-Liquid Phase Equilibrium in Solutions of Alkanethiol-Coated Nanoparticles

**Ezequiel R. Soulé\*, Cristina E. Hoppe, Julio Borrajo, and Roberto J. J. Williams\***

*Institute of Materials Science and Technology (INTEMA), University of Mar del Plata and National Research Council (CONICET), J. B. Justo 4302, 7600 Mar del Plata, Argentina*

\* Authors to whom correspondence should be addressed. Tel.: 54-223-4816600. Fax: 54-223-4810046. E-mail addresses: ersoule@fi.mdp.edu.ar, williams@fi.mdp.edu.ar.

ABSTRACT. A thermodynamic model for a mixture of alkanethiol-coated nanoparticles (NPs) and low molecular weight (non-polymeric) solvent is developed, and calculations of liquid-liquid phase equilibrium for different values of NP core radius, alkanethiol chain length, solvent molar volume and alkanethiol-solvent interaction parameter, are presented. The model takes into account the swelling of the organic coronas and the dispersion of particles with swollen coronas in the solvent. The energetic interaction between alkyl chains and solvent is considered, both within the corona and between the outer alkyl segments and free solvent. Swelling involves mixing of alkanethiol chains and solvent in the corona and stretching of the organic chains. Dispersion considers an entropic contribution based on Carnahan – Starling equation of state and an enthalpic term calculated considering the surface contacts between alkyl segments placed in the external boundary of the corona and the molecules of free solvent. Two different kinds of phase equilibrium are found. One of them, observed at high values of the interaction parameter, is the typical liquid-liquid equilibrium for compact NPs in a poor solvent where a complete phase separation is observed when cooling (increasing the interaction parameter). The second liquid-liquid equilibrium is observed at low values of the interaction parameter, where swelling of coronas is favored. In this region two different phases co-exist, one more concentrated in NPs that exhibit relatively compact coronas and the other one more diluted in NPs with extended coronas. In diluted solutions of NPs the deswelling of the fully extended coronas takes place abruptly in a very small temperature range, leading to a solution of compact NPs. This critical transition might find practical applications similar to those found for the abrupt shrinkage of hydrogels at a critical temperature.

**Introduction**

Rational combination of nanoparticles (NPs) and polymers opens new ways to the development of advanced materials with applications in optoelectronic, sensing, catalysis, magnetic recording and several other fields.[1] In almost any case, real success of these materials strongly depends on the ability to control aggregation and spatial distribution of the particles in the matrix.[2-4] Common strategies to achieve dispersion of NPs in a polymeric host involve some kind of surface functionalization that can be attained through weak interactions (van der Waals, π-stacking), ionic interactions or covalent bonding.[4]

Many synthetic procedures leading to high quality NPs involve surface coverage with *n*-alkyl chains functionalized with an organic group that is bonded to the surface of the NP such as thiols, amines or phosphine oxides.[3] In many cases, simply mixing of a polymer with the alkyl-coated NPs can conduct to uncontrolled aggregation. Nevertheless, it has been demonstrated that high stability and self assembly capacity of these NPs can be exploited for the design of materials with useful applications and unique morphologies.[5-7] Vossmeyer et al.[8] demonstrated that dodecyl-coated gold NPs can be assembled with dendrimers to give composite films with applications in chemical vapor sensing. Alkyl-coated gold and silica nanocrystals have been organized in high-molecular weight block copolymer films to give hierarchically ordered multicomponent materials.[9] Some recent examples have shown that controlled alkyl-coated NPs aggregation can also be used to design materials with new and unexpected properties. Phase separation in the presence of an electric field, of blends formed by alkyl-coated CdSe nanorods and polystyrene (PS), has been exploited for the design of hexagonal arrays of crystals that can find use in photovoltaic devices.[10] Materials with well differenced magnetic behavior have been obtained by changing solely the spatial distribution of oleic acid-coated maghemite NPs in polymeric matrices.[11] Fractal aggregates formed by dodecyl-coated Au NPs on the surface of a hybrid matrix have been obtained by hydrolysis and polycondensation of bridged silsesquioxane precursors.[5] Nano and micron-sized gold colloidal crystals have been assembled in epoxy-based physical gels by polymerization induced phase separation of a mixture of the monomers with the individual NPs.[6] Proper selection of the monomers, alkyl chain length and polymerization conditions have been demonstrated to be of paramount importance for the development of the desired morphology through phase separation and assembly of NPs in the matrix. Some studies have focused on analyzing the effects of temperature and pressure,[12] and the nature of the coating chain[13,14] on phase behavior and self-assembly.

Understanding the thermodynamic principles that lead to phase separation and self-assembly can be very useful in designing new materials, and the ability of predicting phase behavior can help to avoid a trial-and-error procedure that in most cases is used. Despite all the experimental work on alkyl-coated NPs previously described, most of the theoretical work on phase separation and self-assembly in NPs solutions or blends, refer to non-coated NPs. Monte-Carlo simulations,[15,16] dissipative particle dynamics,[17] molecular dynamics,[18] PRISM theory,[19] Density Functional theory,[20] are some of the tools and models used. Ginzburg[21] developed a simple thermodynamic model for mixtures of hard-sphere NPs and flexible polymers, based on a combination of Flory Huggins theory and Carnahan-Starling equation of state. This model gives an explicit expression for the free energy, as a function of composition. Luedtke and Landman[22] performed MD simulations for alkyl-coated NPs, and they observed self-assembly of the NPs in body-centred cubic and face-centred-cubic lattices.

Among the thermodynamic models used in the literature to analyze NP solutions, the one developed by Ginzburg for NP – polymer blends is probably the most simple, and is capable of predicting liquid – liquid phase separation. We have used this model to analyze theoretically polymerization-induced phase separation in mixtures by means of constructing ternary phase diagrams for a monomer – polymer – nanoparticle mixture.[23] In this model, hard spherical nanoparticles are taken into account. To our knowledge, a simple model for alkyl-coated nanoparticles that takes into account the nature of the organic corona has not been developed.

In this paper we build an expression for the free energy of a mixture of a non-polymeric solvent and alkyl-coated nanoparticles, based on Ginzburg´s model. A refinement of this model is employed in the interaction term to take into account that the number of nearest neighbours of a molecule is proportional to its surface area.[24] This enables to take into account the disparity in size of both components of the solution. Besides, a term is added to the free energy, to take into account swelling of the organic coronas. This term is based in expressions derived for micelles by Leibler et al.[25] Liquid – liquid phase equilibrium is calculated for several values of molecular parameters (core radius, alkyl-chain length, solvent molar volume and interaction parameter), in the range of experimental interest.

**Model**

When alkyl-coated particles are mixed with a solvent, the organic coronas may swell to a certain degree, and the particles with swollen coronas are then dispersed in the remaining free solvent. It is necessary then to distinguish between the global volume fractions of coated nanoparticles ($\phi_p$) and solvent ($\phi_s = 1 - \phi_p$), defined as the respective volume fractions in the initial mixture, the volume fraction of free solvent ($\phi_s^{free}$) and particles with swollen coronas ($\phi_p^{swell} = 1 - \phi_s^{free}$) after the swelling process, and the volume fraction of solvent ($\phi_s^c$) and alkyl chains ($\phi_a^c = 1 - \phi_s^c$) within the corona.

The degree of swelling can be characterized by $L$, the length of the alkyl chains when the corona is swollen. The limiting values of $L$ are the value corresponding to the chains fully extended, $L_{ext}$, and the value corresponding to the chains fully compacted (no swelling), $L_{min}$, which can be calculated considering that at $L_{min}$ the alkyl chains fill all the volume of the corona:

$$N_{cp} v_{chain} = \frac{4}{3}\pi \left[ \left( R_p + L_{min} \right)^3 - R_p^3 \right] \quad (1)$$

where $v_{chain}$ is the molecular volume associated to the alkyl chain in a compact phase as will be discussed later, and is proportional to the number of carbon atoms in the chain, $n$. $R_p$ is the particle radius and $N_{cp}$ is the number of chains per particle, which can be calculated from the particle radius and the grafting area of a chain, $s$, as $N_{cp} = 4\pi R_p^2/s$. In a lattice composed of cells with arbitrary volume $v_{ref}$, the number of cells occupied by a chain can be calculated as $r_c = v_{chain}/v_{ref.}$. A scheme of particles exhibiting coronas with different degrees of swelling is shown in Figure 1.

When the corona is swollen, the volume of solvent present in the corona is the difference between the volume of the swollen corona and the volume of the compact corona:

$$V_s^{cor} = \frac{4}{3}\pi\left[(R_p+L)^3 - (R_p+L_{min})^3\right] \qquad (2)$$

It has been found experimentally[26] that not all the volume of the corona is accessible for the solvent, typically the first two segments of the alkyl chains cannot be reached by solvent molecules. The volume of the corona effective for alkyl chains – solvent mixing is:

$$v_{cor} = N_{cp}r_c v_{ref} f + \frac{4}{3}\pi\left[(R_p+L)^3 - (R_p+L_{min})^3\right] \qquad (3)$$

The first term is the volume of the alkyl chains accessible for the solvent and the second term is the volume of solvent. The factor $f$ takes into account that only a portion of each chain is available for mixing. This factor is estimated as $(n-2)/n$. The volume fraction of solvent in the corona can be calculated as

$$\phi_s^c = \frac{V_s^{cor}}{v_{cor}} \qquad (4)$$

The volume fractions of particles with swollen coronas and of free solvent can be calculated as:

$$\phi_p^{swell} = \frac{V_p^{swell}}{V} = \frac{n_p v_p^{compact}\left(\frac{R_p+L}{R_p+L_{min}}\right)^3}{V} = \phi_p\left(\frac{R_p+L}{R_p+L_{min}}\right)^3; \quad \phi_s^{free} = 1 - \phi_p^{swell} \qquad (5)$$

where $V_p^{swell}$ is the volume of particles with swollen coronas and $V$ the total volume of the mixture; $v_p^{compact}$ is the volume of one compact particle.

We will consider the total Gibbs free energy as composed by two contributions, one due to the swelling of the alkyl coronas, and the other one due to the dispersion of the particles with swollen coronas in the free solvent. Following the ideas developed for micelles,[25] the swelling free energy of a single particle's corona can be approximated by the sum of two contributions, a mixing free energy (due to the mixing of alkyl chains and solvent), and an elastic free energy (due to stretching of alkyl chains):

$$g^{swell} = \Delta g^{mix} + g^{el} \qquad (6)$$

The first term can be calculated from the Flory-Huggins theory:

$$\Delta g^{mix} = kT \frac{v_{cor}}{v_{ref}} \left[ \frac{\phi_s^c}{r_s} \ln(\phi_s^c) + \chi \phi_s^c \phi_a^{*c} \right] \qquad (7)$$

where $k$ is Boltzmann constant, $T$ is temperature, and $\phi_a^{*c}$ is an effective alkyl volume fraction for interactions in the corona (the calculation of this variable is detailed in the Appendix), $r_s$ is the number of cells occupied by a solvent molecule (= $v_s/v_{ref}$, where $v_s$ is the volume of a solvent molecule), and $\chi$ is the interaction parameter. As mentioned before, due to the high grafting density of alkyl chains the solvent does not get into contact with the surface of the hard core and with the first two –CH$_2$- segments of the alkyl chains. Therefore, the interaction parameters considers contacts between the solvent and the organic chains. The factor $v_{cor}/v_{ref}$ is the number of cells occupied by the corona, and the factor into brackets is the mixing free energy per cell (only the entropic contribution due to the solvent is considered, because, as the alkyl chains are fixed, their translational entropy vanishes[25]).

The elastic free energy is calculated according to the theory of rubber elasticity:[27]

$$g^{el} = kT \frac{1}{2} N_{cp} \left[ \left( \frac{L}{L_{min}} \right)^2 + 2 \frac{L_{min}}{L} - 3 \right] \qquad (8)$$

In addition to swelling, there is a free energy contribution due to the dispersion of the particles with swollen coronas in the free solvent, as mentioned before. This free energy has entropic and enthalpic contributions,

$$\Delta G^{MIX} = -T\Delta S^{MIX} + \Delta H^{MIX} \qquad (9)$$

The entropic contribution can be approximated as:

$$\Delta S^{MIX} = -k\left( n_s^{free} \ln\left(\phi_s^{free}\right) + n_p \ln\left(\phi_p^{swell}\right) + n_p \frac{4\phi_p^{HS} - 3\left(\phi_p^{HS}\right)^2}{\left(1-\phi_p^{HS}\right)^2} \right) \qquad (10)$$

where $n_s^{free}$ is the number of molecules of free solvent, $n_p$ the number of particles, and $\phi_p^{HS}$ a hard-sphere particle volume fraction. This expression is based on the model by Ginzburg[21,23] for solutions of spherical nanoparticles in (polymeric) solvents. In this model particles are considered as hard spheres, and the configurational entropy is calculated from Carnahan – Starling equation of state. In order to account for the fact that the particles have a hard core and a soft corona, we use an effective hard sphere volume fraction in the Carnahan-Starling correction (last term in equation 10). This effective volume fraction is calculated taking into account the experimentally[28] and theoretically[22] observed fact that when the particles self-assembly in a FCC lattice, the center-to-center distance is $2R_p+L_{ext}$. The hard sphere radius is taken then as $R_p+L_{ext}/2$, and the hard sphere volume fraction is calculated as $\phi_p^{HS} = \phi_p(R_p+L_{ext}/2)^3/(R_p+L_{min})^3$.

The enthalpic term $\Delta H^{MIX}$ arises from energetic interactions between the particles with swollen coronas and the free solvent. Van der Waals long-range interactions between particles cores are not considered based in the molecular dynamics results that show that for particles where the alkyl chain length is comparable to the core radius about 99% of the total interaction energy between particles is due to the interaction of alkyl chains.[22] Van der Waals interactions will become important when the size of the core is much larger than the length of the organic chains. The enthalpic term can be calculated considering the surface contacts between alkyl segments placed in the external boundary of the corona and the molecules of free solvent:

$$\Delta H^{MIX} = kT\chi_a \frac{C_p}{z} n_p \varphi_s^{free} \qquad (11)$$

where $C_p$ is the number of contacts of outer alkyl segments with free solvent, $z$ the cell coordination number and $\varphi_s^{free}$ the area fraction of free solvent (it represents the probability of outer alkyl segments being in contact with free solvent). This expression enables to take into account correctly the interaction energy between organic groups located at the external boundary of the soft corona and the solvent. The interaction parameter $\chi_a$ is defined per unit area ($C_p$ is proportional to the area), and it is different from $\chi$, the interaction parameter per unit volume. The relationship between both is $\chi_a = \chi A_{chain}^{1/2}/4v_{ref}^{1/3}$ (see Appendix), where $A_{chain}$ is the transversal area of an alkyl chain. $C_p$ is calculated as:

$$\frac{C_p(L)}{z} = \frac{4}{3}\pi R_p^2 \frac{1}{v_{ref}^{2/3}} \left[ \frac{\left(1+\frac{L_{min}}{R_p}\right)^2 - \frac{A_{chain}}{s}}{L_{ext} - L_{min}}(L_{ext} - L) + \frac{A_{chain}}{s} \right] \qquad (12)$$

Details on the calculation of these terms are presented in the Appendix.

Assuming that the solvent molecules are spherical, we can calculate the area fraction of free solvent as:

$$\varphi_s^{free} = \frac{n_s^{free} A_s}{n_s^{free} A_s + n_p A_p^{swell}} = \frac{n_s^{free} R_s^2}{n_s^{free} R_s^2 + n_p (R_p + L)^2} = \frac{\dfrac{\phi_s^{free}}{R_s}}{\dfrac{\phi_s^{free}}{R_s} + \dfrac{\phi_p^{swell}}{R_p + L}} \quad (13)$$

where $A_s$ and $A_p^{swell}$ are the superficial areas of a solvent molecule and a particle with a swollen corona, $R_s = [3v_s/(4\pi)]^{1/3}$ is the radius of a solvent molecule.

The total free energy of the mixture is the sum of the free energy of swelling the coronas of the population of nanoparticles, $n_p g^{swell}$, and the free energy of dispersing the nanoparticles with swollen coronas, $G^{MIX}$. This leads to the following dimensionless expression:

$$G^* = \frac{\Delta G}{MkT} = \frac{\phi_p}{r_p} g^{swell} + \frac{\phi_s^{free}}{r_s} \ln(\phi_s^{free}) + \frac{\phi_p}{r_p} \left( \ln(\phi_p^{swell}) + \frac{4\phi_p^{HS} - 3(\phi_p^{HS})^2}{(1-\phi_p^{HS})^2} \right) + \chi_a \frac{C_p}{z} \frac{\phi_p}{r_p} \varphi_s^{free}$$
(14)

where $M$ is the number of cells and $r_p = 4/3\pi(R_p + L_{min})^3/v_{ref}$ is the number of cells occupied by a particle with completely compacted chains.

For a homogeneous solution with a global composition $\phi_s$ and a specified value of the interaction parameter, the equilibrium degree of swelling will be given by the value of $L$ that minimizes the total free energy of the solution in the range $L_{min} \leq L \leq L_{ext}$:

$$L^{eq}(\phi_s, \chi) \leftarrow \min_L \left[ G^*(L, \phi_s, \chi) \right] \quad (15)$$

If the curve $G^*(L^{eq}, \phi_s, \chi)$ vs. $\phi_s$ has a change of curvature, phase separation exists, where the global composition of each phase is determined by the double-tangent rule, and the value of $L^{eq}$ can in principle be different in each phase (because the composition is different).

In order to solve the model, it is necessary to fix the values of all the molecular parameters. The length of the fully extended alkanethiol chains is calculated as a function of the number of carbon atoms:[29]

$$L_{ext} = 3.17 + 1.27(n-1) \text{ Å} \tag{16}$$

The grafting area of the chains is $s = 21 \text{Å}^2$.[22] This value implies that the molecules are packed with a density higher than the liquid bulk value, and close to the solid density, as has been noted in previous works.[22,28] In this work the density used is such that $v_{chain} = 21\text{Å}^2 L_{ext}$. Also it was considered that $A_{chain} = s$. The reference volume enters the model only as a scale factor that determines the specific value of the interaction parameter. The selected reference volume is 26.67 Å$^3$ equal to the volume of a -CH$_2$- segment of the alkyl chain. In order to represent typical experimental systems, the particle radius was varied between 2 and 6 nm (20 and 60 Å), $n$ between 6 and 18, and $v_s$ between 50 and 200 cm$^3$/mol (83 and 332 Å$^3$/molecule).

**Results and Discussion**

Figure 2 shows the dimensionless free energy as a function of solvent volume fraction, for $R_p = 2$ nm, $n = 18$ and $v_s = 100$ cm$^3$/mol, for several values of interaction parameters. For $\chi = 0$, the curvature is always positive, so the solution is homogeneous in the whole composition range. For $\chi = 0.45$, there is a change in curvature at high concentrations of solvent indicating the presence of a liquid-liquid equilibrium (shown with a dashed line). But when the interaction parameter is further increased to $\chi = 1$, this change in curvature disappears and the solution becomes homogeneous again. By increasing further the value of the interaction parameter, a new liquid-liquid equilibrium is found (dashed line) as can be seen from the curvature of the free energy function for $\chi = 2.8$. So, a sequence of states going from homogeneous to phase separated to homogeneous and again to phase separated is predicted when increasing the interaction parameter.

Equilibrium phase diagrams are presented in terms of $\chi^{-1}$, which can be considered as a dimensionless temperature for an upper critical solution temperature (UCST) behavior, as a function of $\phi_s$. They were calculated for different combinations of the varied parameters, $R_p$, $n$ and $v_s$ and are presented in Figures 3, 4 and 5. Figure 3 is parametric in $n$ ($R_p = 2$ nm, $v_s = 100$ cm$^3$/mol and $n = 6$, 12 and 18), Figure 4 in $R_p$ ($n = 18$, $v_s = 100$ cm$^3$/mol and $R_p = 2$, 4 and 6 nm), and Figure 5 in $v_s$ ($n = 18$, $R_p = 2$ nm, and $v_s = 50$, 100 and 200 cm$^3$/mol).

For every system except for $R_p = 2$ nm, $v_s = 100$ cm$^3$/mol, and $n = 6$, two different regions of phase equilibrium exist as was inferred from Figure 2. At high values of $\chi$ (low $\chi^{-1}$), the typical liquid-liquid equilibrium (from now on, referred as equilibrium 1), was found; at low values of $\chi$ (high $\chi^{-1}$), another phase equilibrium region (equilibrium 2), was present as a closed loop. This second L-L phase equilibrium is not predicted by models that consider the NPs as hard spheres. As will be analyzed later its appearance is associated with the presence of the soft shell composed of organic chains.

The equilibrium 1 region of the phase diagram exhibits a critical point located at low solvent volume fractions. However, the critical point predicted with the unmodified Ginzburg's model lies in the region of high solvent volume fractions.[21,23] This different behavior arises from the fact that in the present model the interactions are proportional to the areas of the components. The area fraction of the solvent (small spheres) is significantly larger than its volume fraction due to the small contribution of the area fraction of the NPs (large spheres). This leads to a shift of the maximum of $\Delta H^{MIX}$ vs. $\phi_s$ to the region of high concentrations of NPs. This is the reason of the shift of the critical point to the region of low $\phi_s$ values. A similar behavior is observed when using other thermodynamic models like UNIQUAC that consider interactions to be proportional to the contact area. When applied to solutions of spherical molecules the critical point predicted by UNIQUAC is close to the pure component with largest size.

In the equilibrium 1 region of the phase diagram both phases located along a horizontal tie-line have different concentrations of nanoparticles. Calculations show that in both phases the organic coronas are almost completely compacted exhibiting no noticeable swelling. Increasing the value of the interaction parameter leads to the trivial case of no significant solubility of NPs in the poor solvent (tie-lines join the two pure components: compact NPs and the pure solvent; as shown by the curve for $n = 6$ in Figure 3 every curve exhibits a sharp curvature close to the axis of pure solvent getting asymptotic to this axis). It is necessary to point out that we are considering only liquid-liquid phase separation (disordered phases); if crystallization of NPs is allowed a crystal-liquid equilibrium should also be calculated.

On the other hand, equilibrium 2 takes place in conditions of low $\chi$ values, where swelling is favored. Calculations show that the degree of swelling of the coronas is different in both phases. The influence of different parameters on both liquid-liquid equilibria will be now discussed starting by the equilibrium 1 region of the phase diagrams.

Figure 3 shows that increasing the size of the alkyl chain ($n$) favors miscibility in the equilibrium 1 region. As mentioned before, the coronas are completely compacted in this case, so the swelling free energy plays no role in this equilibrium. Miscibility is determined by the competition between the entropy and the enthalpy of mixing. As can be seen in equation (14), the dependence of the entropic and enthalpic terms with $n$ is through $r_p$, $\phi_p^{HS}$, $C_p(L)/z$, and $\varphi_s^{free}$. By examining the dependence of every one of these factors on the size of the alkyl chains it is concluded that increasing $n$ produces a decrease in both the enthalpic contribution (favoring mixing) and the entropic contribution (favoring demixing). The net result shown by numerical calculations is an increase in miscibility in the equilibrium 1 region when increasing the length of the alkyl chain.

Figure 4 shows the effect of modifying the core radius, $R_p$. In this case there is a crossover of curves in the equilibrium 1 region: at high concentrations of solvent the increase in $R_p$ favors miscibility, but at low concentrations of solvent, miscibility is favored by decreasing $R_p$. At high concentrations of solvent the entropic contribution provided by NPs is very small (due to the low value of their volume fraction) and the enthalpic contribution is directly proportional to $(R_p+L_{min})^{-1}$ (the dependence of $\varphi_s^{free}$ on $R_p$ is not significant). Therefore, the decrease in the enthalpic contribution by increasing $R_p$ produces an increase in miscibility. At very low concentrations of solvent, $\varphi_s^{free}$ is proportional to $R_p+L_{min}$, so the enthalpy is not a function of $R_p+L_{min}$ and the entropic term contributed by NPs determines the influence of $R_p$ on miscibility. An increase of $R_p$ decreases the absolute value of the entropic contribution to free energy favoring demixing. This is due to the increase in both $r_p$ and in the Carnahan-Starling hard-sphere correction.

It is interesting to observe that an increase in both the length of the alkyl chains ($n$) and the size of the core ($R_p$) has a similar effect in diluted solutions of NPs. Both effects increase miscibility by the decrease in the surface per unit volume exposed to the poor solvent (decrease of the enthalpic contribution). Again, this effect arises by the fact that in the present model the interactions are proportional to the areas of the components. The model predicts that largest particles are more miscible with the solvent, a fact that might be questioned on the basis that fractionation of NPs by precipitation invariably separates fractions in order of descending NPs size (largest NPs are precipitated first).[30,31] Fractionation is performed by decreasing the solvent quality by addition of a non-solvent followed by centrifugation. Separation is performed by a gravitational field that produces fractionation of NPs in order of descending size, independently of any thermodynamic consideration.

The effect of modifying the molar volume of the solvent is shown in Figure 5. In this case, as expected, miscibility decreases monotonically with increasing solvent size. The decrease in the entropic contribution of the solvent is responsible for this trivial effect.

It is interesting to note that the value of interaction parameter (temperature) at which phase separation occurs is not significantly affected by the global composition in the region of high solvent concentration. This means that, according to the thermodynamic model, any solution of nanoparticles, regardless how diluted it is, will phase separate at a temperature that is principally determined by the molecular parameters ($R_p$, $n$, $v_s$), and not by the concentration of NPs (assuming that the interaction parameter does not depend on composition). A small increase in the interaction parameter (decrease in temperature) when surpassing the equilibrium curve leads to almost pure phases. This means that, when a solution of nanoparticles is cooled, the phases will rapidly purify as temperature decreases. This can be better seen in Figure 6, which is a plot of the phase boundaries for $R_p = 2$ nm and $n = 18$, in the region of $\phi_s > 0.95$.

Let us now turn our attention to the equilibrium 2 region. As mentioned previously, this equilibrium takes place in conditions where swelling of coronas is favored. Calculations show that in this region the coronas have different values of $L_{eq}$ in both phases. In order to better understand this phenomenon, Figure 7 shows $L_{eq}/v_{ref}^{1/3}$ as a function of composition, for different values of $\chi$, in a particular system ($R_p = 2$ nm, $n = 18$, $v_s=100$ cm$^3$/mol). For these particular values of the parameters, the equilibrium 2 region is the largest closed loop shown in both Figures 3 and 4. The curves for $\chi$: -3, -0.7 and 0 are located in the miscible region of the phase diagram plotted in both Figures. The curves for $\chi$: 0.375, 0.55 and 0.65 intercept the equilibrium 2 region for particular values of the solvent concentration. The curve for $\chi = 0.75$ is located slightly below the equilibrium 2 region.

The dashed line in Figure 7 represents the maximum theoretical value of $L$ when all the available solvent swells the coronas, $\phi_s^{free} = 0$ for $L \leq L_{ext}$. Free solvent appears only after the coronas have been swollen to their maximum extent ($L_{ext}$). For $\chi = -3$ the calculated curve lies close to the asymptotic theoretical curve. The added solvent is mostly used to swell the coronas leaving a very small amount of free solvent until $L_{ext} =$ is attained. For $\chi = -0.7$ and 0 the degree of swelling is significantly smaller than the theoretical limit meaning that the entropic contribution of the free solvent in the blend with NPs with swollen coronas contributes significantly to the decrease of the free energy of the mixture. Therefore, the solvent distributes inside the coronas and outside the NPs as free solvent. An increase in $\chi$ produces a corresponding increase in the solvent volume fraction at which $L_{ext}$ is reached. For $\chi$: 0.375, 0.55 and 0.65 curves exhibit inflexion points (as for $\chi = 0.375$) or discontinuities (as for $\chi= 0.55$ and 0.65). For this range of $\chi$ values, $G^*(L^{eq}, \phi_s, \chi)$ vs $\phi_s$ exhibits a change of curvature, as shown in Figure 2, indicating that phase separation is predicted. Equilibrium compositions determined by the double-tangent rule are those intercepting the equilibrium 2 region of Figures 3 and 4. These compositions determine different values of $L^{eq}$ in both phases (read from any particular curve of Figure 7 with a $\chi$ value located in the two-phase region). The left branch at high NPs concentration has coronas that are less swollen (lower value of $L^{eq}$) than the right branch located at low NPs concentration where coronas are swollen to the maximum extent, $L_{ext}$. For $\chi = 0.75$ the coronas are almost fully compacted in all the concentration range (Figure 7). This system is a homogeneous solution of almost compact NPs in a poor solvent.

Although no details are presented here, there are two critical points located at the boundary of the closed loop. One of these points is located in the pure solvent axis and both points are joined by two equilibrium curves defining the boundary of the closed loop. These two equilibrium curves represent solutions with different composition and with NPs exhibiting different swelling of their coronas: more compact in the branch of high concentration of NPs and extended in the branch of low concentration of NPs.

Phase separation in the two-phase region can be better understood by imagining vertical trajectories going down in the phase diagram plotted in Figure 3. These trajectories represent an increase in the interaction parameter or a corresponding temperature decrease for a typical $\chi \sim T^{-1}$ behavior. Depending on the initial concentration of NPs these cooling trajectories may or may not intercept the equilibrium 2 region. Let us first assume a trajectory located at the left of the equilibrium 2 region so that there is no interception when cooling. The initial homogeneous solution may be located in a concentration range where coronas of NPs are swollen to the maximum extent, $L_{ext}$. At a certain temperature in the cooling process coronas will begin to contract; $L^{eq}$ decreases continuously as shown by vertical trajectories in Figure 7 following an increase in the interaction parameter. During cooling coronas expel the solvent and become fully contracted. But these compact NPs still form a homogeneous solution in a poor solvent until the equilibrium 1 region is attained. Cooling in this region produces a complete phase separation of the compact NPs and the pure solvent.

In cooling trajectories that intercept the equilibrium 2 region, the deswelling of coronas takes place simultaneously with a phase separation process. One of the phases has relatively compact coronas while the other phase exhibits completely expanded coronas. The system leaves this two-phase region as a homogeneous solution of NPs with compact coronas dispersed in a poor solvent.

Figure 8 illustrates this behavior, showing curves of degree of phase segregation, $f_V$ (defined as the ratio of volume of segregated phase with respect to total volume), and linear extension of the coronas of both phases for a system with $R_p = 2$ nm, $n = 18$, and $v_s = 100$ cm$^3$/mol, across the equilibrium 2 region. Two different global concentration of solvent are considered: 0.7 (Figure 8a) and 0.9 (Figure 8b).

When the system is located to the left of the equilibrium 2 critical point (Figure 8a), the coronas in mother phase have an intermediate degree of swelling, and the phase segregated when equilibrium 2 is reached has fully swollen coronas. With further cooling, the amount of segregated phase first increases and then decreases, while the degree of swelling in the mother phase decreases. When the system leaves equilibrium 2, the system is again homogeneous and the coronas are almost fully compacted.

If the starting solution is located to the right of the equilibrium 2 critical point (Figure 8b) the cooling trajectory must be interpreted in terms of the discontinuous curves shown in Figure 7. In this case, coronas remain fully extended during the cooling process until the right branch of the equilibrium 2 region is attained. A differential cooling from this point generates a differential amount of a new phase more concentrated in NPs with compact coronas. The mother phase becomes more diluted and continues to keep extended coronas. With further cooling the amount of segregated phase with compact coronas keeps increasing, it becomes predominant and finally a temperature is attained at which the original phase disappears and a homogeneous solution of compact NPs in a poor solvent is obtained.

When the initial solution becomes more diluted in NPs the temperature range where the transition from extended to compact coronas takes place becomes very small. This can be seen in Figure 8b, where the interval of $\chi^{-1}$ for coexistence of phases is narrow and it gets narrower for increasing solvent concentrations. In much diluted solutions of NPs the existence of a critical temperature (in fact a very small temperature range) where coronas change from the fully extended to the compact condition is predicted. This is a similar process than the shrinkage of hydrogels observed at a critical temperature, a phenomenon that finds several practical applications.

As the equilibrium 2 region expresses the swelling-deswelling behavior of the coronas of NPs, the extension of this region increases when the size of coronas increases with respect to the size of the core as can be observed in Figures 3 and 4. As shown in Figure 5, increasing the solvent size shifts the equilibrium to higher values of $\chi^{-1}$ (coronas deswell at higher temperatures).

**Conclusions**

A thermodynamic model describing the liquid-liquid phase equilibrium of alkanethiol-coated nanoparticles in a solvent has been developed. This model takes into account the swelling of the organic coronas and the dispersion of particles with swollen coronas in the solvent. The energetic interaction between alkyl chains and solvent has been considered, both within the corona and between the outer alkyl segments and free solvent. It was found that two regions of liquid-liquid phase equilibrium exist.

At high values of the interaction parameter (low temperatures) the typical liquid-liquid equilibrium between two phases of different concentration of nanoparticles is observed, with coronas fully compacted in both phases. Increasing the interaction parameter (cooling) produces the complete phase separation of compact NPs from the solvent. In diluted solutions, the miscibility of compact NPs in a poor solvent increases when increasing their size either by increasing the size of the core or the corona. This is due to the decrease of the surface interactions per unit volume between the corona and the solvent. An original finding was the prediction of a liquid-liquid phase separation at relatively low values of the interaction parameter (high temperatures). In this region two different phases co-exist, one more concentrated in NPs that exhibits relatively compact coronas and the other one more diluted in NPs with extended coronas. In diluted solutions of NPs the deswelling of the fully extended coronas takes place abruptly at a critical temperature (in fact in a very small temperature range), leading to a solution of compact NPs. In our opinion this theoretical prediction is one of the most relevant results of the present model. The design of NPs stabilized by organic chains than can undergo this critical transition at convenient solvent/temperature combinations might be used in drug-delivery applications with the drug initially loaded in the swollen corona. A similar sharp transition between swollen and shrunk states has been found for several types of hydrogels used in biological and medical applications.[32] This transition takes place when the hydrogel enters the immiscibility region of its phase diagram with the solvent, either through an increase in temperature (lower critical solution temperature behavior, LCST) or a decrease in temperature (upper critical solution temperature behavior, UCST). Similar transitions have been recently predicted for polymer brushes,[33] which are the equivalent of the soft corona extended over a planar surface. Certainly, the existence of a similar critical transition in NPs grafted by organic chains, predicted by the thermodynamic model, still needs experimental verification.

Although the set of numerical results was illustrated for the case of NPs stabilized with alkanethiol chains, the existence of two liquid-liquid equilibrium regions and critical transitions can be present in a variety of NPs stabilized with organic chains of different chemical structure and length.

**Acknowledgements**

This paper is dedicated to Prof. Hugo de Lasa on the occasion of his 65$^{th}$ anniversary. The financial support of the University of Mar del Plata, the National Research Council (CONICET), and the National Agency for the Promotion of Science and Technology (ANPCyT), is gratefully acknowledged.

**Appendix**. **Calculation of the Parameters Appearing in the Interactions Terms**

The parameter $C_p$ for a swollen particle is estimated by linear interpolation between two limits: compact corona and fully extended corona. For a compact corona, the whole surface of the particle is formed by alkyl segments, so $C_p$ is the coordination number of a solid sphere in the lattice. The number of cells placed in the surface of the sphere can be calculated as the ratio of the area of the sphere and the area of a cell (taken as $v_{ref}^{2/3}$). These cells have neighbors inside and outside the sphere, being the latter the relevant ones for determining the coordination number of the sphere. The fraction of outside neighbors will depend on the lattice geometry, if we assume that it is 1/3 of the cell coordination number, then $C_p(L_{min})/z = 4\pi(R_p+L_{min})^2/3v_{ref}^{2/3}$ (this expression is the same that Ginzburg used in his model).[21]

For a fully extended corona, there are $N_{cp}$ chain segments on the surface, occupying a total area of $N_{cp} A_{chain}$, where $A_{chain}$ is the transversal area of an alkyl chain. Considering again 1/3 of the cell coordination number, $C_p(L_{ext})/z = N_{cp}A_{chain}/3v_{ref}^{2/3} = 4\pi R_p^2 A_{chain}/(3sv_{ref}^{2/3})$.

Interpolating,

$$\frac{C_p(L)}{z} = \frac{4}{3}\pi R_p^2 \frac{1}{v_{ref}^{2/3}} \left[ \frac{\left(1+\frac{L_{min}}{R_p}\right)^2 - \frac{A_{chain}}{s}}{L_{ext}-L_{min}}(L_{ext}-L) + \frac{A_{chain}}{s} \right] \quad (A1)$$

If $n_1$ molecules of component "1" interact with component "2", Flory-Huggins expression for the interaction free energy is $\Delta H/kT = \chi n_1 r_1 P = Bn_1v_1P$, where $r_1$ is the number of cells occupied by a molecule of component 1, $P$ is the probability of finding a cell occupied by component "2", and $B=\chi/v_{ref}$, is a dimensionless interaction parameter per unit volume. In this theory, the interaction energy is proportional to the volume of each component, because they are considered as "linear" molecules, so that the number of contacts is proportional to the molecular volume. But in the case where a component is a solid (e.g. a solid sphere), the interaction energy is proportional to the area (as can be seen in the previous expression, where $C_p(L)$ is proportional to $R_p^2$). In order to calculate an interaction parameter per unit area we will assume that alkyl chains of component 1 are composed of a string of $n$ cubes with side equal to $A_{chain}^{1/2}$ (so that the cross-sectional area is $A_{chain}$). The external area of such a molecule is $a_1 = (4/6)n(6A_{chain}) = 4nA_{chain}$, as 4 of the 6 faces of each cube (total area of $6A_{chain}$) are external and 2 are internal (neglecting end effects). The volume of this molecule is $v_1 = nA_{chain}^{3/2}$. A characteristic length of this molecule may be defined as $l_1 = v_1/a_1 = A_{chain}^{1/2}/4$. The Flory-Huggins interaction energy can be written as $\Delta H/kT = (\chi/v_{ref})n_1 l_1 a_1 P = (\chi_a/v_{ref}^{2/3})n_1 a_1 P$, where $\chi_a$ is an interaction parameter defined per unit external area. Then,

$$\chi_a = \chi \frac{l_1}{v_{ref}^{1/3}} \qquad (A2)$$

The effective alkyl volume fraction for interactions in the corona, $\phi_a^{*c}$, is calculated considering only the contacts between alkyl chains and solvent *inside* the corona. The total number of alkyl cells available for interaction with the solvent is $N_{cp}r_c f$, so that the total number of cells adjacent to an alkyl cell is $N_{cp}r_c f z$. But as we discussed previously, a fraction of the alkyl cells is placed in the outer surface of the corona and interacts partially with the free solvent and not with the solvent in the corona. So the total number of contacts between alkyl cells and solvent inside the corona is $(N_{cp}r_c fz - C_p(L))\phi_s^c$. According to this, the effective volume fraction for interactions in the corona is:

$$\phi_a^{*\,c} = (N_{cp}r_c f - \frac{C_p(L)}{z})\frac{v_{ref}}{V_{cor}} = \phi_a^c - \frac{C_p(L)}{z}\frac{v_{ref}}{V_{corona}} \tag{A3}$$

**Legend to the Figures**

**Figure 1.** Scheme of alkanethiol-coated nanoparticles in a solvent; (a) with an intermediate degree of swelling; (b) fully shrunk (left) and fully swollen (right) cases.

**Figure 2.** Dimensionless free energy as a function of solvent volume fraction for different values of $\chi$: 0, 0.45, 1 and 2.8, and $R_p$= 2 nm, $n = 18$ and $v_s = 100$ cm$^3$/mol. Dashed lines show double-tangent points indicating equilibrium between two phases.

**Figure 3.** Phase diagrams calculated with $R_p = 2$ nm, $v_s = 100$ cm$^3$/mol, and $n = 6$ (dot), 12 (dash), and 18 (solid).

**Figure 4.** Phase diagrams calculated with $n = 18$, $v_s = 100$ cm$^3$/mol, and $R_p = 2$ (dot), 4 (dash), and 6 nm (solid).

**Figure 5.** Phase diagrams calculated with $R_p = 2$ nm, $n = 18$, $v_s = 200$, (dot), 100 (dash), and 50 cm$^3$/mol (solid).

**Figure 6.** Phase separation values of the interaction parameter calculated with $R_p = 2$ nm, $n = 18$, $v_s = 200$, (dot), 100 (dash), and 50 cm$^3$/mol (solid), in diluted solutions of NPs.

**Figure 7.** Equilibrium degree of swelling as a function of composition for $R_p = 2$ nm, $n = 18$, $v_s = 100$ cm$^3$/mol and several values of $\chi$: -3, -0.7, 0, 0.375, 0.55, 0.65, 0.75. The dashed line represent the maximum value of $L$ when all the available solvent swells the coronas, $\phi_s^{free} = 0$ for $L \leq L_{ext}$.

**Figure 8.** Fraction of phase with swollen coronas (upper) and linear extension of coronas of both phases (lower) as a function of the inverse of the interaction parameter, for a system with $R_p = 2$ nm, $n = 18$, and $v_s = 100$ cm$^3$/mol, and two different global solvent volume fractions: 0.7 (a) and 0.9 (b).

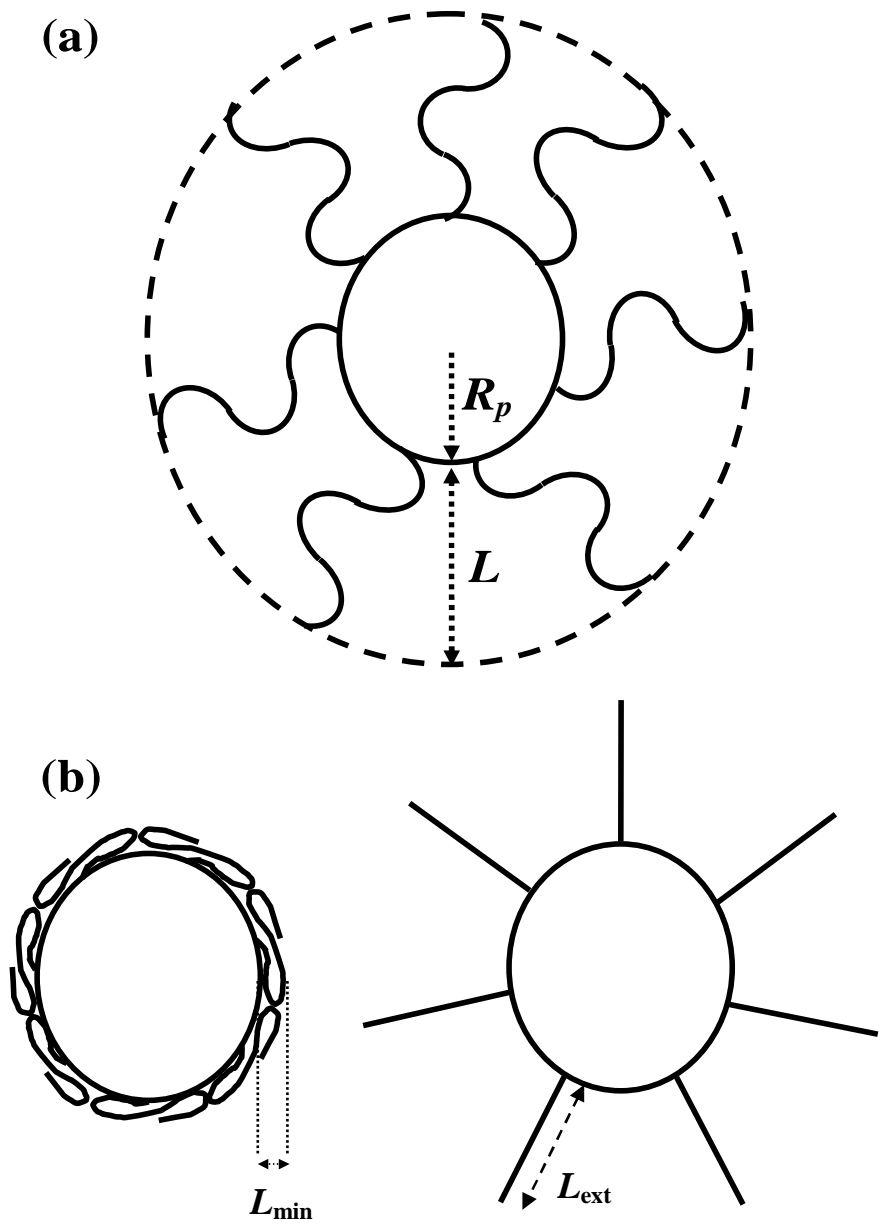

**Figure 1**

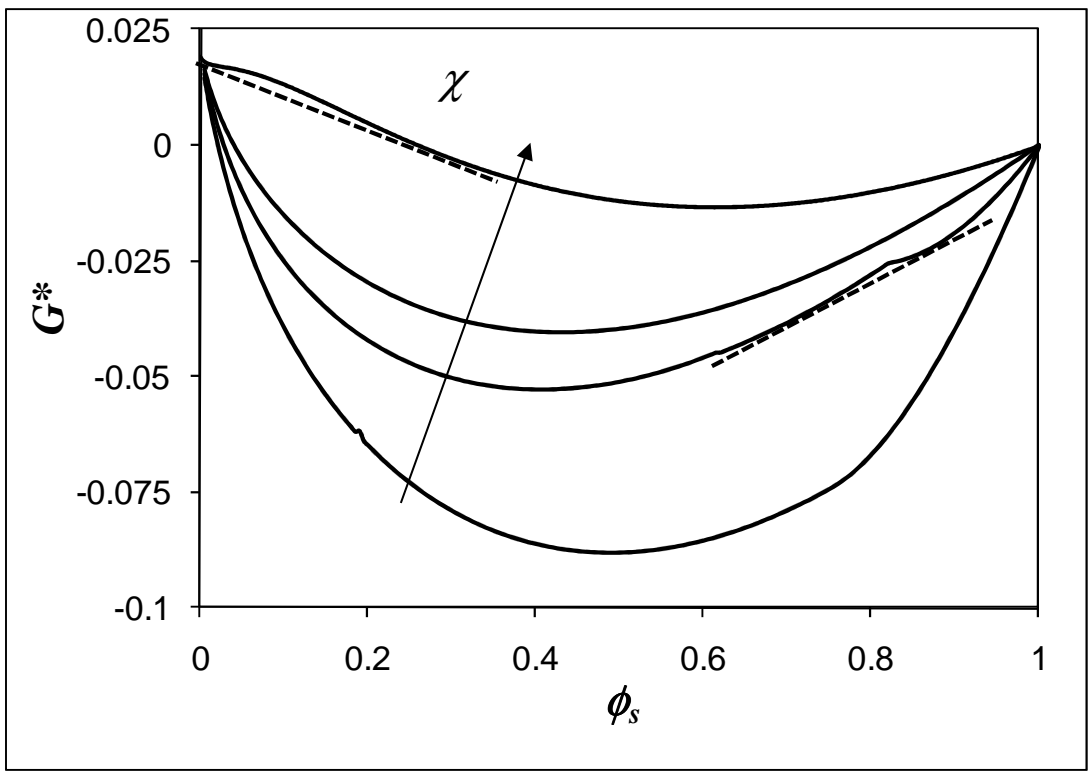

**Figure 2**

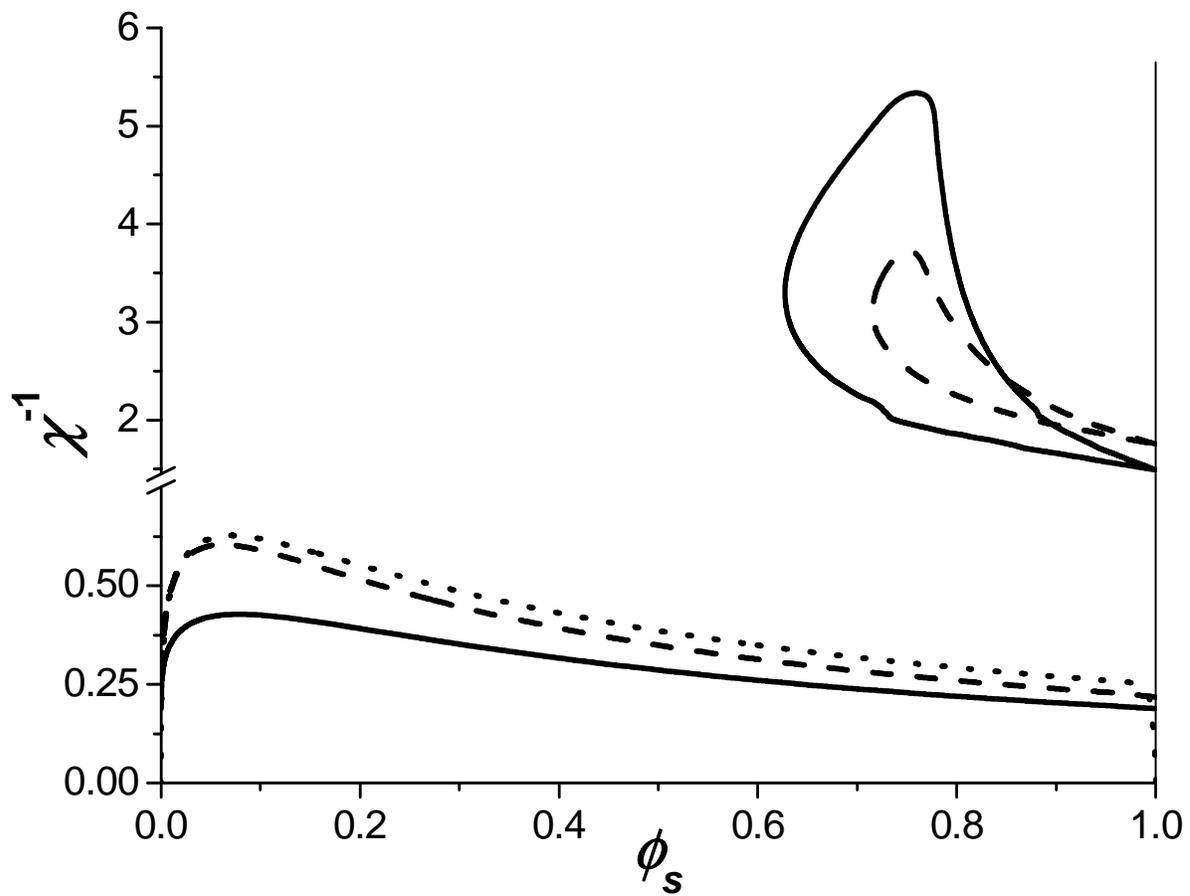

**Figure 3**

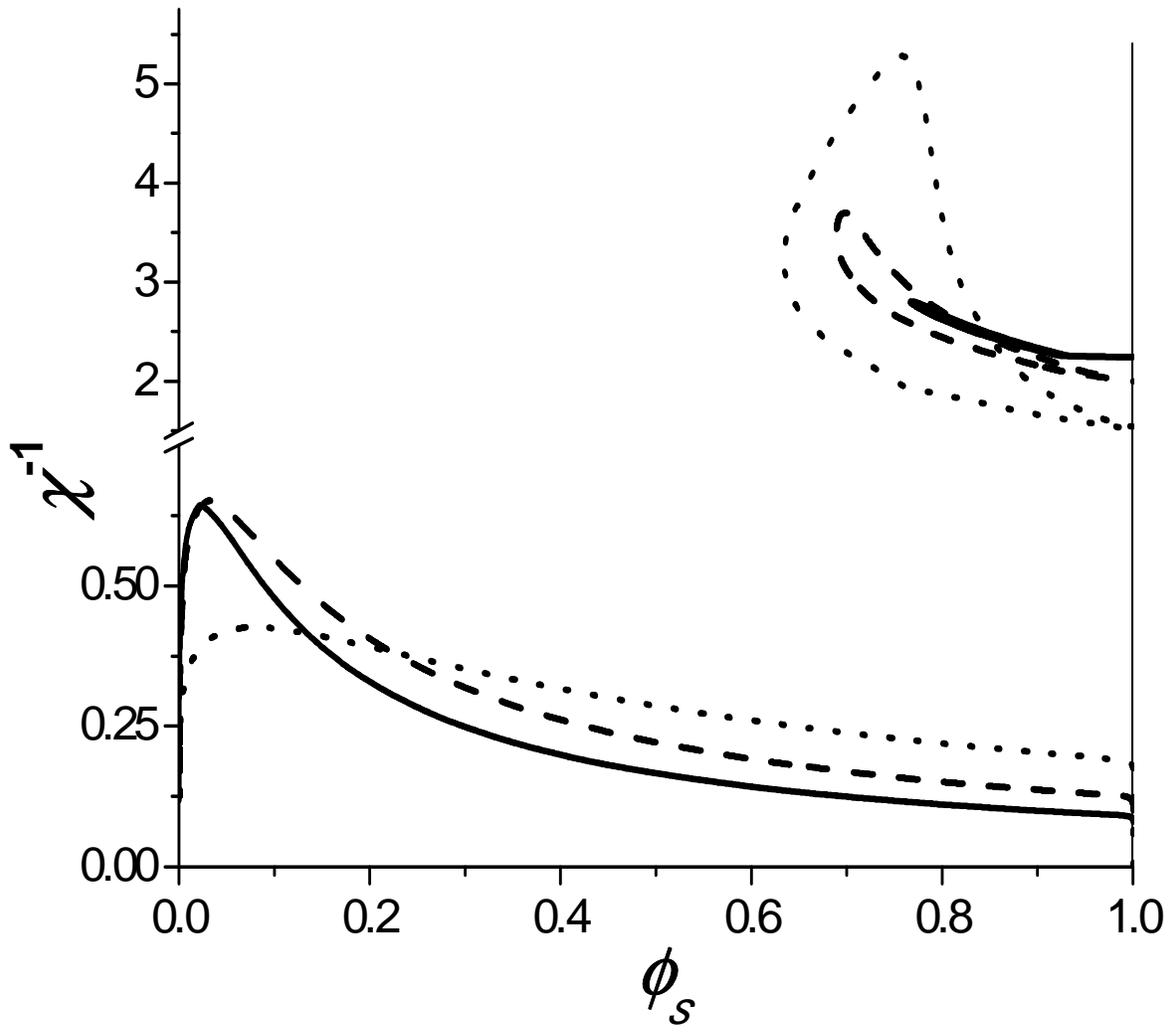

**Figure 4**

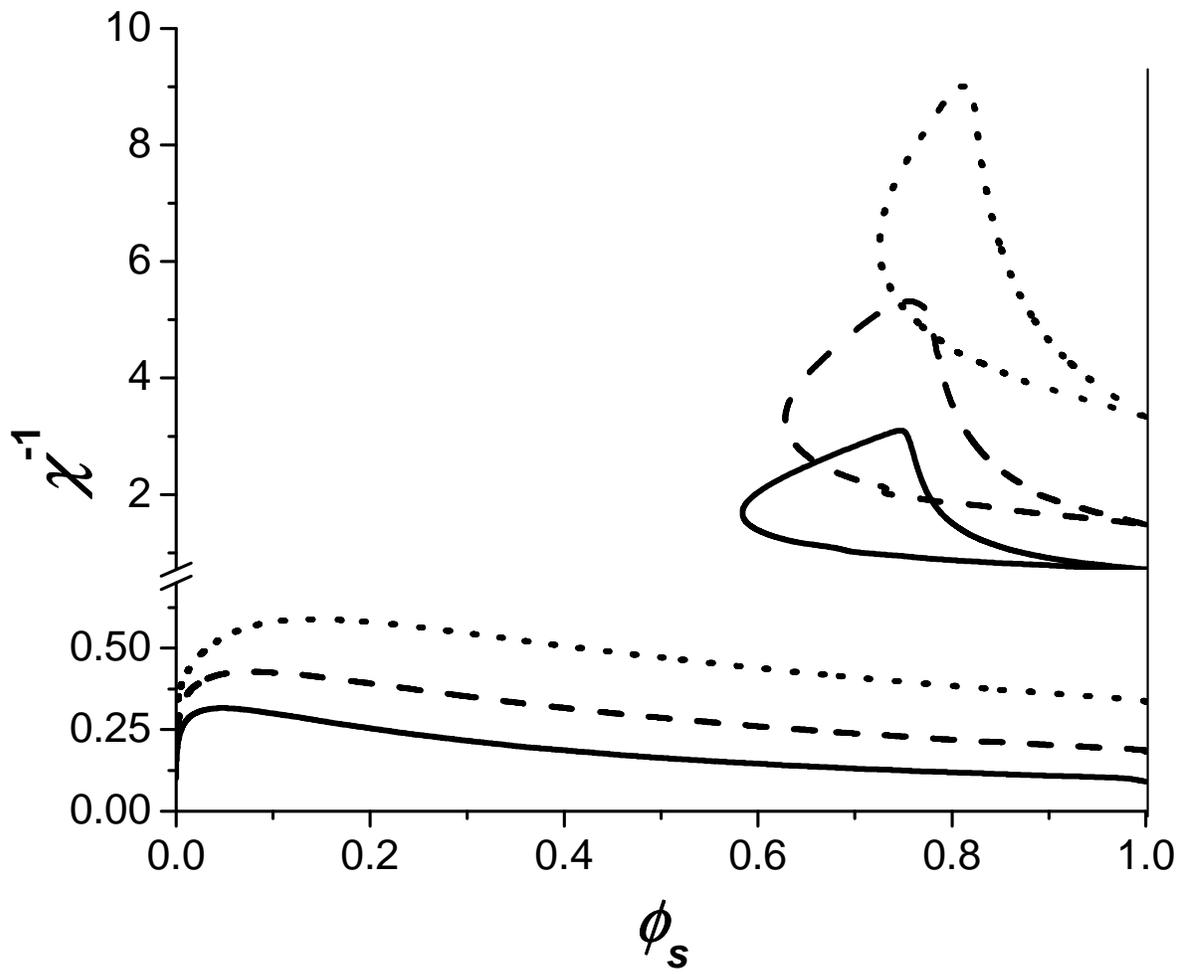

**Figure 5**

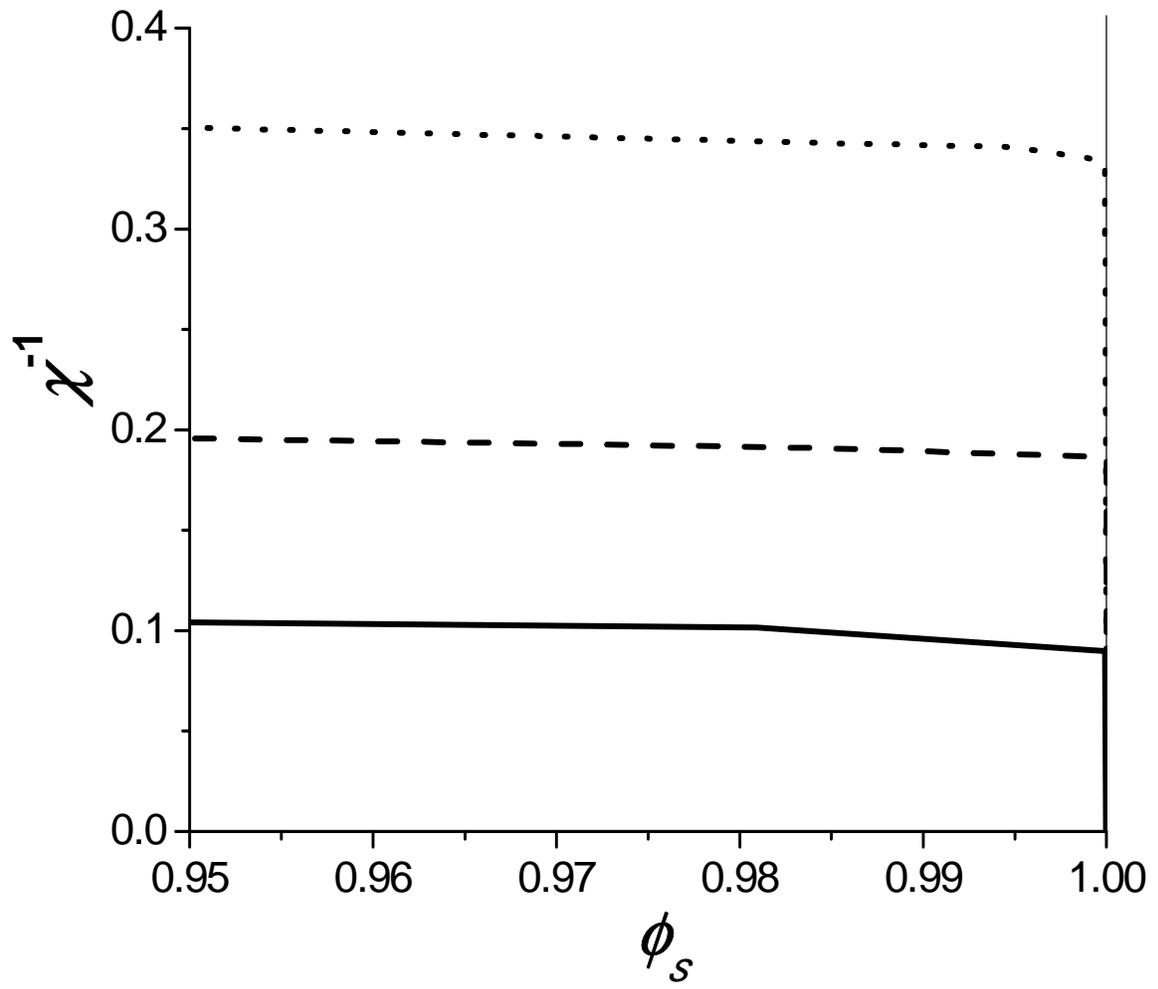

**Figure 6**

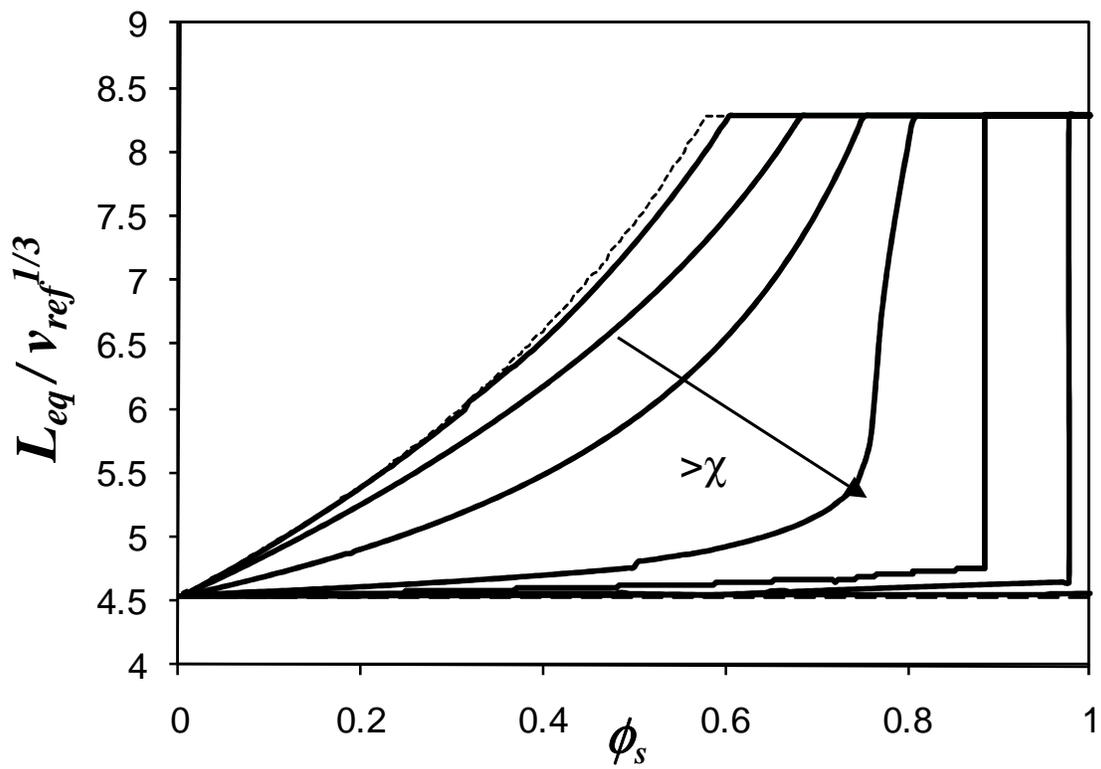

**Figure 7**

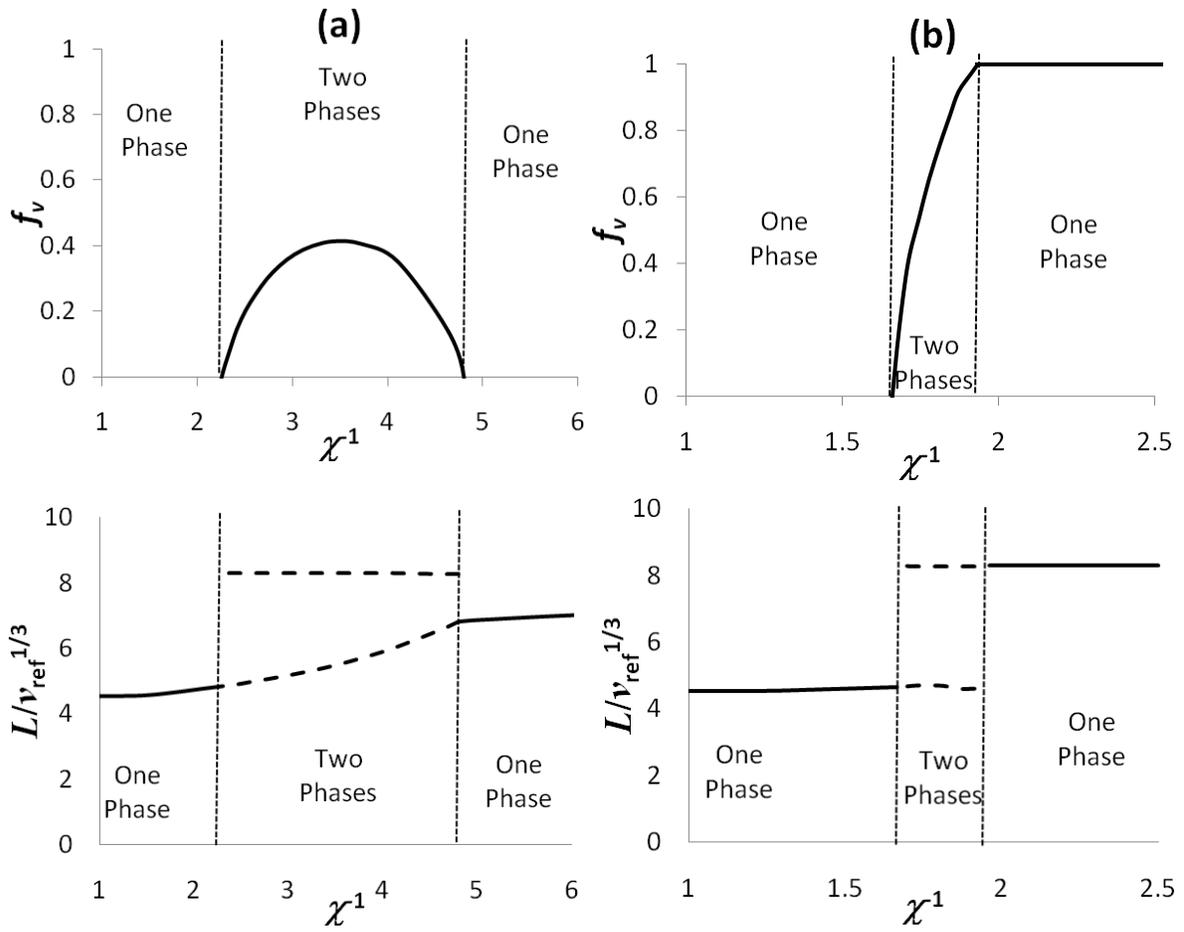

**Figure 8**